\documentclass[twocolumn,aps,showpacs,superscriptaddress]{revtex4}
\usepackage{graphicx}
\usepackage{amsmath}
\usepackage{bm}
\begin{document}
\title{Power-law velocity distributions in granular gases}
\author{E.~Ben-Naim}
\email{ebn@lanl.gov}
\affiliation{Theoretical Division and Center for Nonlinear Studies,
Los Alamos National Laboratory, Los Alamos, New Mexico 87545}
\author{B.~Machta}
\email{benjamin_machta@brown.edu}
\affiliation{Department of Physics, Brown University,
Providence, Rhode Island 02912}
\affiliation{Department of Physics, University of Massachusetts,
Amherst, Massachusetts 01003}
\author{J.~Machta}
\email{machta@physics.umass.edu}
\affiliation{Department of Physics, University of Massachusetts,
Amherst, Massachusetts 01003}
\begin{abstract}
We report a general class of steady and transient states of granular
gases. We find that the kinetic theory of inelastic gases admits
stationary solutions with a power-law velocity distribution, $f(v)\sim
v^{-\sigma}$.  The exponent $\sigma$ is found analytically and depends
on the spatial dimension, the degree of inelasticity, and the
homogeneity degree of the collision rate. Driven steady-states, with
the same power-law tail and a cut-off can be maintained by
injecting energy at a large velocity scale, which then cascades to
smaller velocities where it is dissipated.  Associated with these
steady-states are freely cooling time-dependent states for which the
cut-off decreases and the velocity distribution is self-similar.
\end{abstract}

\pacs{45.70.Mg, 47.70.Nd, 05.40.-a, 81.05.Rm}

\maketitle

\section{Introduction}

The statistical physics of granular gases is unusual in many ways
\cite{bp,pl,pb}. Shaking a box of beads, no matter how hard, fails to
generate a thermal distribution of energy. Instead, the velocity
distributions are not Maxwellian \cite{lcdkg,rm,ao} and energy may be
distributed unevenly in space \cite{dlk,kwg,gzb} or among different
components of a polydisperse granular media \cite{wp,fm}.  Moreover,
spatial correlations may spontaneously develop \cite{bcdr}. Granular
gases also exhibit interesting collective phenomena such as shocks
\cite{rbss,slk}, clustering \cite{my,ou,lh,nbc,mwl}, and hydrodynamic
instabilities \cite{gz,km}. Energy dissipation, which results from
inelastic collisions, is largely responsible for this rich
phenomenology.

Dilute granular matter can be studied systematically using kinetic
theory. This approach has been used to quantitatively model situations
where the dynamics are primarily collisional
\cite{jr,bdks,bdl,ig}. Kinetic theory has been used to derive
transport coefficients in the continuum theory of rapid granular
flows, and it has also been used to model freely evolving and driven
granular gases.

Spatially homogeneous systems are a natural starting point for
investigations of granular gases. Theoretical, computational, and
experimental studies show that the system cools indefinitely without
energy injection, and that it reaches a steady-state when energy is
injected to counter the dissipation.  In the freely cooling case, the
velocity distribution follows a self-similar form and in the forced
case, the velocity distribution approaches a steady-state.  In either
case, the velocity distributions have sharp tails, and in particular,
all of their moments are finite.

In this study, we consider the very same spatially homogeneous
granular gases and show that there is an additional family of steady
and transient states. First, we demonstrate that for a special,
analytically soluble case, the unforced Boltzmann equation admits
non-trivial stationary states where the velocity distribution has a
power-law high-energy tail. Then, we show that in general, the tail of
the distribution obeys a linear equation and use this master equation
to demonstrate that stationary states with power-law tails are
generic, existing for arbitrary dimension and arbitrary collision
rules. The characteristic exponents are obtained analytically
\cite{bm} .

The mechanism responsible for these stationary states is an energy
cascade from large velocity scales to small velocity scales that
occurs due to the inelastic particle collisions. Driven steady-states
with the same characteristic exponent and a high velocity cut-off can
be maintained by injecting energy at a large velocity scale to
compensate for the energy dissipated in the cascade.  We confirm these
steady-states using Monte Carlo simulations. We propose that such
steady states can be experimentally realized in driven granular
systems in which energy is injected at large velocities.

There is also a family of closely related freely cooling
time-dependent states.  We demonstrate this explicitly in
one-dimension.  In these transient states, the velocity distribution
coincides with the stationary distribution up to some large velocity
scale, but falls off exponentially beyond that scale. This cut-off
velocity obeys Haff's cooling law and decreases algebraically with
time until the power-law range collapses. The velocity distribution is
self-similar and the underlying scaling function is obtained
analytically using the linear Boltzmann equation. These freely cooling
states are confirmed using numerical integration of the Boltzmann
equation.

This paper is organized as follows. The system is set-up in section II
and a special case is solved in section III. Dynamics of large
velocities and the linear Boltzmann equation are described in section
IV. Stationary states are detailed in section V, driven steady-states
in section VI and transient states in section VII. We conclude in
section VIII.

\section{Inelastic gases}

We study a spatially homogeneous system of identical particles
undergoing inelastic collisions. First, we consider one-dimension
where the linear collision rule is
\begin{equation}
\label{rule-1d}
v_{1,2}=pu_{1,2}+qu_{2,1}
\end{equation}
with $v_{1,2}$ the post collision velocity and $u_{1,2}$ the
pre-collision velocity. The collision parameters $p$ and $q$ obey
$p+q=1$. The relative velocity is reduced by the restitution
coefficient $r=1-2p$ as follows: $(v_1-v_2)=-r(u_1-u_2)$.  In each
collision, momentum is conserved, but the total kinetic energy
decreases.  The energy loss is $\Delta E=pq(u_1-u_2)^2$. Energy
dissipation is maximal for the extreme case of completely
inelastic collisions $(r=0,p=1/2)$ and it vanishes for the extreme
case of elastic collisions $(r=1,p=0)$.

In this study, we consider the general collision rate
\begin{equation}
\label{rate-1d}
K(u_1,u_2)=|u_1-u_2|^\lambda
\end{equation}
with $0\leq \lambda\leq 1$ the homogeneity index. For particles
interacting via the central potential $U(r)\sim r^{-\nu}$, the
homogeneity index is $\lambda=1-2\frac{d-1}{\nu}$ \cite{rd,sk}.  There
are two limiting cases: (i) Hard-spheres, where the collision rate is
linear in the velocity difference, $\lambda=1$, are used to model
ordinary granular media; (ii) Maxwell-molecules, where the collision
rate is independent of the velocity are used to model granular media
with certain dipole interactions \cite{max,true,e,avb,ia}.

Let $f(v,t)$ be the distribution of particles with velocity $v$ at
time $t$. It is normalized to unity, $\int dv f(v)=1$ (henceforth the
dependence on $t$ is left implicit). For freely evolving and spatially
homogeneous systems the distribution obeys the Boltzmann equation
\begin{eqnarray}
\label{be-1d}
\frac{\partial f(v)}{\partial t}=\iint du_1 du_2 |u_1-u_2|^\lambda f(u_1)f(u_2)
\qquad\\
\qquad\qquad\times\Big[\delta(v-pu_1-qu_2)-\delta(v-u_1)\Big].\nonumber
\end{eqnarray}
In this master equation, the kernel equals the collision rate
 (\ref{rate-1d}) and the gain and loss terms simply reflect the
collision law  (\ref{rule-1d}). The Boltzmann equation assumes perfect
mixing as the probability of finding two particles at the same
position is taken as proportional to the product of the individual
particle probabilities. It is exact when the strong condition of
perfect mixing or ``molecular chaos'' is met, but it is only
approximate when the particle positions are correlated.

One well-known solution of the this equation is the ``homogeneous
cooling state'' where the velocity distribution is self-similar in the
long time limit \cite{ep,bds},
\begin{equation}
\label{cooling}
f(v,t)\simeq \frac{1}{v_0}\psi\left(\frac{v}{v_0}\right)
\end{equation}
with the characteristic velocity $v_0$.  Applying dimensional analysis,
the collision rate $K\propto v_0^\lambda$ should be inversely
proportional to time, $K\sim t^{-1}$. This leads to Haff's cooling law
\cite{pkh}
\begin{equation}
\label{haff}
v_0\sim t^{-1/\lambda}.
\end{equation}
Alternatively, it follows from the rate equation \hbox{$dv_0/dt\propto
  -v_0^{1+\lambda}$}, implying that exponential decay occurs for the
limiting case of Maxwell molecules. Statistics of energetic
particles are characterized by the tail of the distribution, and
for freely cooling states, there is a stretched exponential decay
\cite{ep,eb03,bbrtv}
\begin{equation}
\label{psi}
\psi(z)\sim
\exp\left(-|z|^\lambda\right),
\end{equation}
for $\lambda>0$ as $|z|\to\infty$.

In the freely cooling states, all energy is dissipated from the system
and the particles come to rest, $f(v,t)\to \delta(v)$ as
$t\to\infty$. Thus, the system reaches a trivial stationary state. Are
there any nontrivial stationary states? Quite surprisingly, the answer
is yes.  Our main result is that generically, there is a family of
nontrivial stationary solutions of the Boltzmann equation.

\section{An exact solution}

The stationary velocity distribution can be obtained analytically for
one-dimensional Maxwell molecules. Since the governing equation
(\ref{be-1d}) is in a convolution form, it is natural to employ the
Fourier transform \cite{krupp}, $F(k)=\int dv\, e^{ikv} f(v)$.  The
stationary state ($\partial/\partial t\equiv 0$) satisfies the
non-local and non-linear equation \cite{bk,bblr}
\begin{equation}
F(k)=F(pk)F(qk).
\end{equation}
Normalization implies $F(0)=1$.

For elastic collisions, $p=0$, every distribution is a stationary state,
but this is a one-dimensional anomaly, because in higher dimensions,
the stationary distribution is always Maxwellian \cite{max}. For all
$0\leq p\leq 1$ and $p+q=1$, there is a family of stationary
solutions
\begin{equation}
F(k)=\exp\left(-v_0|k|\right),
\end{equation}
characterized by the arbitrary typical velocity $v_0$. Performing
the inverse Fourier transform, the velocity distribution is a
Lorentz (Cauchy) distribution \cite{bs}
\begin{equation}
f(v)=\frac{1}{\pi v_0}\frac{1}{1+(v/v_0)^2}.
\end{equation}
This distribution decays algebraically at large velocities. For
freely cooling Maxwell-molecules in one-dimension, the velocity
distribution has a related form, a squared Lorentzian \cite{bmp}.

This stationary distribution does not evolve under the collision
dynamics since at each velocity there is perfect balance between
collisional loss and collisional gain. The total energy density and
the total dissipation rate are both divergent due to the shallow tail
of the velocity distribution.

\section{Cascade Dynamics}

To analyze the general behavior, we focus on the dynamics of very
energetic particles.  This allows us to derive the power-law decay and
to obtain the characteristic exponent for all spatial dimensions and
all collision parameters.

\subsection{One-Dimension}

The collision integral in Eq.~(\ref{be-1d}) greatly simplifies in the
limit $v\to\infty$. Since the distribution decays sharply at large
velocities, the product $f(u_1)f(u_2)$ is maximal when one of the
pre-collision velocities is large and the other small. For the gain
term there are two possibilities: either $u_1\gg u_2$ and then
$v=pu_1$ or $u_2\gg u_1$ and then $v=qu_2$. Let us denote the large
velocity by $u$ and the small one by $w$. The double integral
separates into two independent integrals,
\begin{eqnarray}
\frac{\partial f(v)}{\partial t} =\int dw f(w)\int du\,
|u|^\lambda f(u)\qquad\qquad\\
\qquad\qquad\qquad\times[\delta(v-pu)+\delta
(v-qu)-\delta(u)].\nonumber
\end{eqnarray}
Here, the collision rate $|u-w|^\lambda$ was approximated by
$|u|^\lambda$. The integral over the smaller velocity equals one,
and performing the integration over the larger velocity yields
\begin{eqnarray}
\label{lbe-1d} \frac{\partial f(v)}{\partial t}\! =\! |v|^\lambda
\left[\frac{1}{p^{1+\lambda}}f\left(\frac{v}{p}\right)\!+\!
\frac{1}{q^{1+\lambda}}f\left(\frac{v}{q}\right)\!-\!f(v)\right].\quad
\end{eqnarray}
The tail of the velocity distribution satisfies a non-local but {\em
linear} evolution equation.

The linear Boltzmann equation is valid for broader conditions compared
with the full nonlinear Boltzmann equation. The only requirement is
that energetic particles are uncorrelated with slower particles. This
is a weaker condition than the {\it ``stosszahlansatz''} that the two
particle density be equal to a product of one-particle densities.

\begin{figure}[t]
\includegraphics[width=0.2\textwidth]{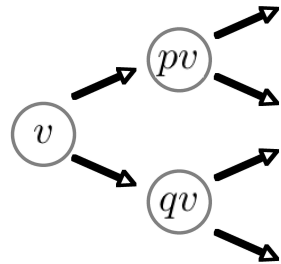}
\caption{The cascade process.} \label{fig:cascade}
\end{figure}

Eq.~(\ref{lbe-1d}) reflects that large velocities undergo the
following cascade process
\begin{equation}
\label{cp-1d} v\to (\,pv\,,\,qv\,),
\end{equation}
with the rate $|v|^\lambda$.  These cascade dynamics follow
directly from the collision rule (\ref{rule-1d}) by setting one of
the incoming velocities to zero. Even though the number of
particles is conserved, the number of energetic particles doubles
in each cascade event (Fig.~\ref{fig:cascade}). Moreover, momentum
is conserved but energy is dissipated in each cascade event: it is
transferred from large velocities to smaller velocities.

\subsection{Arbitrary Dimension}

In general dimensions, the collision rule is
\begin{equation}
\label{rule}
{\bf v}_1={\bf u}_1-(1-p)({\bf u}_1-{\bf
u}_2)\cdot\hat{\bf n}\,\hat{\bf n}.
\end{equation}
Here $\hat{\bf n}\equiv{\bf n}/n$ with $n\equiv |{\bf n}|$ is a unit
vector parallel to the impact direction ${\bf n}$ (connecting the
particle centers), ${\bf v}_{1,2}$ are the post-collision velocities,
and ${\bf u}_{1,2}$ are the pre-collision velocities. The normal (to
$\hat{\bf n}$) component of the relative velocity is reduced by the
restitution coefficient $r=1-2p$ as follows, \hbox{$({\bf v}_1-{\bf
v}_2)\cdot\hat{\bf n} =-r\,({\bf u}_1-{\bf u}_2)\cdot\hat{\bf n}$} and
the energy dissipated equals \hbox{$p(1-p)|({\bf u}_1-{\bf
u}_2)\cdot\hat{\bf n}|^2$}. Similarly, the general collision rate
(\ref{rate-1d}) becomes \hbox{$K({\bf u}_1,{\bf u_2})=|({\bf u}_1-{\bf
u}_2)\cdot\hat{\bf n}|^\lambda$}. The velocity distribution $f_d({\bf
v})$ satisfies
\begin{eqnarray}
\frac{\partial}{\partial t}f_d({\bf v})\!\! &=&\!\!\iiint d\hat{\bf n}\,d{\bf
u}_1\, d{\bf u}_2 |({\bf u}_1-{\bf u}_2)\cdot\hat{\bf n}|^\lambda
f_d({\bf
u}_1)f_d({\bf u}_2)\nonumber\\
&\times&[\delta({\bf v}-{\bf v}_1)-\delta({\bf v}-{\bf u}_1)].
\end{eqnarray}
In addition to integration over the incoming velocities, an additional
integration over the impact direction is required, and this
integration is normalized, $\int d\hat{\bf n}= 1$.  The impact angle
is assumed to be uniformly distributed.

The dynamics of large velocities $v\to\infty$ are simplified as in the
one-dimensional case. The integration over the incoming velocities is
separated into an integral over a small velocity and an integral over
a large velocity ${\bf u}$.  The former integration is immediate,
\begin{eqnarray}
\label{be-a} &&\frac{\partial}{\partial t}f_d({\bf v}) =\iint
d\hat{\bf n}\,d{\bf u} |{\bf u}\cdot\hat{\bf n}|^\lambda f_d({\bf u})\times
\\
&&\left[\delta({\bf v}\!-\!(1\!-\!p){\bf u}\cdot\hat{\bf n}\,\hat{\bf
n})\!+\!\delta({\bf v}\!-\!{\bf u}\!+\!(1\!-\!p){\bf u}\cdot\hat{\bf
n}\,\hat{\bf n})\!+\!\delta({\bf v}\!-\!{\bf
u})\right].\nonumber
\end{eqnarray}
Let $\mu=(\hat{\bf u}\cdot\hat{\bf n})^2$; in other words, if $\theta$
is the angle between the dominant velocity and the impact angle, then
\hbox{$\mu=\cos^2\theta$}. There are two gain terms corresponding to
the two cases \hbox{${\bf v}=(1-p){\bf u}\cdot\hat{\bf n}\,\hat{\bf
n}$} and \hbox{${\bf v}={\bf u}-(1-p){\bf u}\cdot\hat{\bf n}\,\hat{\bf
n}$}. These collision rules, together with the impact angle, dictate
the magnitudes of the post-collision velocity in terms of the
pre-collision velocity, $v=\alpha u$ and $v=\beta u$ with the
following stretching parameters
\begin{subequations}
\label{sp}
\begin{align}
\alpha&=(1-p)\mu^{1/2},\\
\beta&=\left[1-(1-p^2)\mu\right]^{1/2}.
\end{align}
\end{subequations}
The parameter $\alpha$ follows from $\hat{\bf u}=\hat{\bf n}$ and the
parameter $\beta$ is obtained by introducing ${\bf w}={\bf v}-{\bf u}$
and then employing the collision rule \hbox{$w=-{\bf w}\cdot \hat{\bf
n}=(1-p){\bf u}\cdot \hat{\bf n}=(1-p)u\mu^{1/2}$} and the identity
\hbox{$v^2=u^2+w^2-2uw\mu^{1/2}$}.  The integration over the large
velocity ${\bf u}$ includes separate integrations over the velocity
magnitude $u$ and over the velocity direction $\hat{\bf u}$ but since
this angle is a unique function of the impact angle, the latter
integration is immediate. Using the isotropic velocity distribution
\hbox{$f_d({\bf v})\equiv S_d v^{d-1}f(v)$} with \hbox{$S_d\int dv\,
v^{d-1}f(v)=1$} and $S_d$ the area of the $d$-dimensional
unit hypersphere, Eq.~(\ref{be-a}) simplifies to
\begin{eqnarray}
v^{d-1}\frac{\partial f(v)}{\partial t}&=&\iint d\hat{\bf n}\,du\,
|u\mu^{1/2}|^\lambda\,u^{d-1}f(u)\\
&\times&\left[\delta(v-\alpha u)+\delta(v-\beta
u)-\delta(v-u)\right]\nonumber.
\end{eqnarray}
Finally, we simplify the angular integration, \hbox{$d\hat{\bf
n}\propto \sin^{d-2}\theta\, d\theta$}. Denoting the angular
integration with angular brackets $\langle g\rangle \equiv \int
d\hat{\bf n}\,g(\hat{\bf n})$, we have
\begin{equation}
\label{g} \langle g\rangle=C\int_0^1 d\mu\,
g(\mu)\,\mu^{-1/2}(1-\mu)^{\frac{d-3}{2}}.
\end{equation}
The constant $C=1/B(\frac{1}{2},\frac{d-1}{2})$, with $B(a,b)$ the beta
function, is set by normalization. The linear equation governing the
tail of the distribution is therefore
\begin{equation}
\label{lbe} \frac{\partial f(v)}{\partial t}\!=\!
\biggl<\!v^\lambda\mu^{\lambda/2}\!\left(\frac{1}{\alpha^{d+\lambda}}f
\left(\frac{v}{\alpha}\right)\!+\!
\frac{1}{\beta^{d+\lambda}}f
\left(\frac{v}{\beta}\right)\!-\!f(v)\right)\!\biggr>.
\end{equation}

As in one-dimension, large velocities undergo the cascade process
\begin{equation}
\label{cp} v\to (\,\alpha v\,,\,\beta v\,),
\end{equation}
but in general dimension, the stretching parameters acquire a
dependence on the impact angle. In each collision, the total velocity
magnitude increases, despite the fact that the total energy decreases,
as reflected by the following two inequalities
\begin{subequations}
\label{ineq}
\begin{align}
\alpha+\beta&\geq 1,\\
 \qquad \alpha^2+\beta^2&\leq 1.
\end{align}
\end{subequations}
Equalities occur  in the limiting cases: the total velocity
 magnitude is conserved in one dimension where collisions are always
 head-on ($\mu=1$) and, of course, the total energy is conserved for
 elastic collisions ($p=0$). Actually, a stronger statement than
 (\ref{ineq}) holds: the quantity
 $M_s(\alpha,\beta)=\alpha^s+\beta^s-1$ is positive for $s\leq 1$,
 negative for $s\geq 2$, and it may be either positive or negative in
 the range $1<s<2$, depending on the impact angle.

\section{Stationary States}

We have seen that the velocity distribution decays algebraically for
one-dimensional Maxwell molecules. The linear equation for the tail of
the distribution shows that this behavior extends to
all $\lambda$ and all $p$ in one-dimension. The power-law velocity
distribution
\begin{equation}
\label{powerlaw}
f(v)\sim v^{-\sigma}
\end{equation}
satisfies the linear Boltzmann equation (\ref{lbe-1d}) with the time
derivative set to zero when
$p^{\sigma-1-\lambda}+q^{\sigma-\lambda-1}=1$. Since the collision
parameters satisfy $p+q=1$, the characteristic exponent in
one-dimension is simply
\begin{equation}
\label{sigma-1d}
\sigma=2+\lambda.
\end{equation}
Of course, the power-law behavior applies only for the tail of the
distribution.

\begin{figure}[t]
\includegraphics[width=0.45\textwidth]{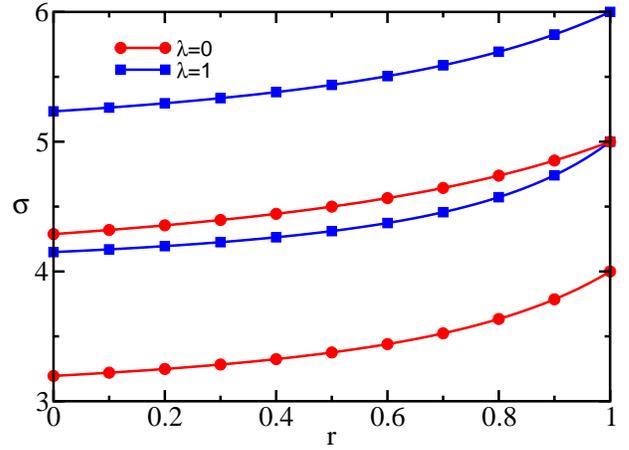}
\caption{The exponent $\sigma$ versus the restitution coefficient $r$
for hard-spheres ($\lambda=1$) and Maxwell molecules ($\lambda=0$).
The top two curves are for $d=3$ and the bottom two curves are for
$d=2$.}
\label{fig:sigma}
\end{figure}

Algebraic behavior holds in arbitrary dimension.  Substituting
Eq.~(\ref{powerlaw}) into the general linear Boltzmann equation
(\ref{lbe}), the characteristic exponent is root of the equation
\begin{equation}
\label{root}
\left<\left(\alpha^{\sigma-d-\lambda}+
\beta^{\sigma-d-\lambda}-1\right)\mu^{\lambda/2}\right>=0.
\end{equation}
This transcendental equation can be re-written explicitly in terms of
the gamma function and the hypergeometric function \cite{gr}
\begin{eqnarray}
\label{main}
\frac{1\!-\!{}_2F_1\!\left(\frac{d+\lambda-\sigma}{2},\frac{\lambda+1}{2},
\frac{d+\lambda}{2},1\!-\!p^2\right)}{(1-p)^{\sigma-d-\lambda}}\!=\!
\frac{\Gamma(\frac{\sigma-d+1}{2})\Gamma(\frac{d+\lambda}{2})}
{\Gamma(\frac{\sigma}{2})\Gamma(\frac{\lambda+1}{2})}.\quad
\end{eqnarray}
The exponent $\sigma\equiv \sigma(d,\lambda,r)$ varies continuously
 with the spatial dimension $d$, the homogeneity index $\lambda$, and
 the restitution coefficient $r$ (figure \ref{fig:sigma}).

According to the bounds (\ref{ineq}) the left hand side of
Eq.~(\ref{root}) is positive when $\sigma-d-\lambda\leq 1$ but
negative when $\sigma-d-\lambda\geq 2$. Therefore, this quantity
changes sign when $1\leq \sigma-d-\lambda\leq 2$ leading to the
relatively tight bounds
\begin{equation}
d+1+\lambda\leq \sigma\leq d+2+\lambda.
\end{equation}
The lower bound (\ref{sigma-1d}) is realized in one-dimension where
the collisions are always head-on, while the upper bound is
approached, $\sigma\to d+2+\lambda$, in the quasi-elastic limit $r\to
1$. We note that the two limiting cases of one-dimension and elastic
collisions do not commute.  Moreover, the zero dissipation limit is
singular: Maxwellian distributions occur when the collisions are
elastic \cite{max}.

Since the energy lost in each collision is proportional to $(\Delta
v)^2$ and the collision rate is proportional to $|\Delta v|^\lambda$,
the energy dissipation rate is related to the following integral,
$\Gamma\sim \bm{\langle} v^{2+\lambda}\bm{\rangle}$ where
$\bm{\langle} g(v)\bm{\rangle} \equiv S_d\int dv
\,v^{d-1}f(v)g(v)$. Hence, the bound $\sigma\leq d+2+\lambda$ implies
that the total dissipation rate is divergent. This is a generic
feature of the stationary solutions, and in fact it shows why Haff's
cooling law $dT/dt=-\Gamma$, where $T=\bm{\langle} v^2\bm{\rangle}$ is
the granular temperature, does not apply: this rate equation assumes
finite dissipation rates.  In contrast, the total energy may be either
finite or infinite because both $\sigma>d+2$ and $\sigma<d+2$ are
possible. The stationary states studied here appear to be
fundamentally different than the infinite energy solutions of the
elastic Boltzmann equation because they require dissipation and
because they always involve infinite dissipation \cite{bc}.

The characteristic exponent increases monotonically with the spatial
dimension, the homogeneity index, and the restitution
coefficient. Thus, fixing $d$ and $\lambda$, the completely inelastic
case ($r=0$) provides a lower bound for $\sigma$ with respect to $r$
(figure \ref{fig:sigma}). For hard-spheres the completely inelastic
limit yields $\sigma=4.1922$ and $\sigma=5.23365$ in two- and
three-dimensions, while for Maxwell molecules the corresponding values
are $\sigma=3.19520$ and $\sigma=4.28807$.

The power-law behavior is in sharp contrast with the stretched
exponential tails $f(v) \sim \exp(-|v|^\delta)$ that typically
characterize granular gases. For freely cooling gases,
$\delta=\lambda$ as in (\ref{psi}), and for thermally forced gases,
$\delta=1+\lambda/2$ \cite{ve,bk03,se,adl}. Both behaviors immediately
follow from the linear Boltzmann equation (\ref{lbe}); in the forced
case, the time derivative in (\ref{lbe}) is replaced by the diffusive
forcing term \hbox{$\partial/\partial t\to D\nabla^2$}. Only in the
limiting case of freely cooling Maxwell molecules do power-law
velocity distributions arise, but the solutions are not stationary and
the characteristic exponent differs from the stationary solutions
\cite{kb,bk02,eb1,eb2}.

\section{Driven Steady-States}

In this section we describe driven, non-equilibrium steady-states
that are identical, except for a high velocity cut-off, to the
stationary states described above.  In these steady-states energy
is injected at a large velocity scale, cascades to small
velocities, and is dissipated over a broad power-law range. The
energy injection scale $V$ must be well separated from the typical
velocity scale $v_0$, but otherwise, the injection mechanism is
not unique.  We study several concrete cases where energy is
injected with a small rate at a large velocity.

As we have seen in the exactly soluble case of one-dimensional Maxwell
molecules, there is a family of steady-state solutions characterized
by the typical velocity $v_0$: if $f(v)$ is a steady-state solution,
so is $v_0^{-d}f(v/v_0)$ for arbitrary $v_0$. Let the energy injection
rate (per particle) be $\gamma$ and let the injection velocity be
$V$. This scale sets an upper cutoff on the velocity distribution,
beyond which the distribution should rapidly vanish. Since the system
is at a steady-state, the dissipation rate
\begin{eqnarray}
 \label{dissipation}
 \Gamma &\sim&  \bm{\langle}v^{2+\lambda}\bm{\rangle} \sim
\int^V dv\, v^{d+1+\lambda}\, v_0^{-d} f(v/v_0)\\
&\sim& V^{\lambda+2}(V/v_0)^{d-\sigma}, \nonumber
\end{eqnarray}
must be balanced by the energy injection rate $\gamma V^2$,
leading to a general relation between the injection rate, the
injection velocity and the typical velocity,
\begin{equation}
\label{bal} \gamma\sim V^\lambda (V/v_0)^{d-\sigma}.
\end{equation}

\begin{figure}[t]
\includegraphics[width=0.45\textwidth]{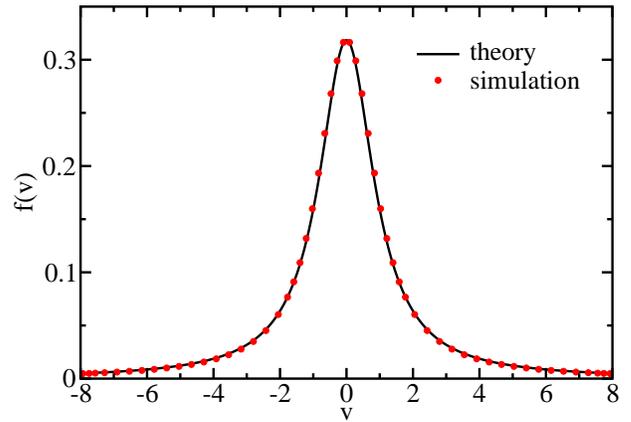}
\caption{The velocity distribution for one-dimensional Maxwell
molecules. The solid line is a Lorentzian and the typical velocity
is $v_0=0.055$.} \label{fig:lorentz}
\end{figure}

To verify the theoretical predictions, we performed Monte Carlo
simulations. Collisions are simulated by selecting two particles at
random with a probability proportional to the collision rate and then
updating their velocities according to the collision rule
(\ref{rule}). Energy is injected with a small rate using the following
``lottery'' implementation. An energy loss counter keeps track of the
cumulative energy loss. With a small rate, a randomly chosen particle
is ``awarded'' an energy equal to the reading on the loss counter.
Subsequently, the loss counter is reset to zero. With this protocol,
the kinetic energy remains practically constant, and moreover, energy
injection occurs only at large velocity scales. For one-dimensional
hard-spheres, we tested a different injection mechanism. The injection
energy was drawn from a Maxwell-Boltzmann distribution with a very
large energy. With a small rate, this energy was added to a randomly
chosen particle.

\begin{figure}[t]
\includegraphics[width=0.45\textwidth]{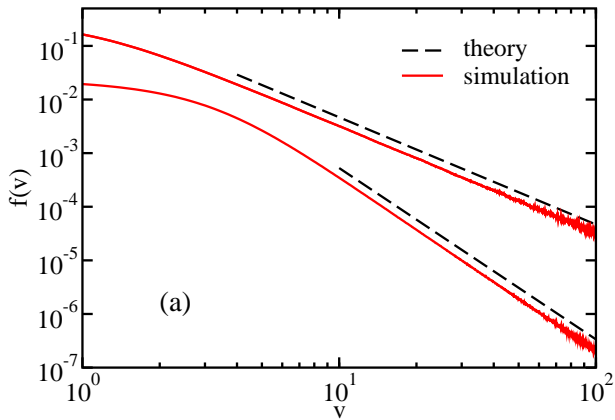}
\caption{The velocity distribution for Maxwell molecules.  The top
curves correspond to one-dimension and the bottom curves to
two-dimensions.}
\label{fig:mm}
\end{figure}

\begin{figure}[t]
\includegraphics[width=0.45\textwidth]{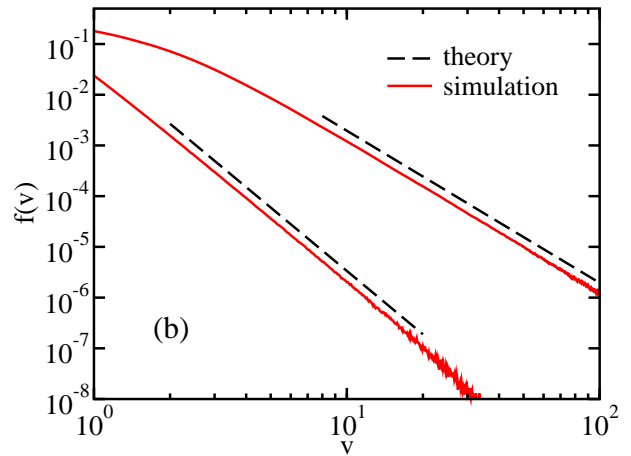}
\caption{The velocity distribution for hard spheres in one-dimension
(top curves) and in two-dimensions (bottom curves).}
\label{fig:hs}
\end{figure}

We simulated completely inelastic Maxwell molecules and hard spheres
in one- and two-dimensions starting with a uniform velocity
distribution with support in the range $[-1:1]$.  After a short
transient, the system reaches a steady-state. For the special case of
one-dimensional Maxwell molecules, we verified that the velocity
distribution is Lorentzian (figure \ref{fig:lorentz}). In all cases,
the tail of the velocity distribution decays as a power-law, and the
exponent $\sigma$ is in excellent agreement with the theoretical
prediction, Eq.~(\ref{main}).  Maxwell molecule simulation results are
displayed in figure \ref{fig:mm} and hard sphere simulation results in
figure \ref{fig:hs}.

The energy balance relation (\ref{bal}), combined with the constant
energy condition $\bm{\langle} v^2\bm{\rangle}\sim 1$, imposed in our
simulations, yields an estimate for the typical velocity.  Different
behaviors emerge for finite energy and infinite energy distributions.

When $\sigma<d+2$, the constant energy constraint implies
$V^{d+2-\sigma}\sim v_0^{d-\sigma}$, that combined with energy
balance (\ref{bal}) reveals how the maximal velocity and the
typical velocity scale with the injection rate
\begin{subequations}
\begin{align}
V&\sim \gamma ^{-\frac{1}{2-\lambda}}, \\
v_0&\sim\gamma^{\frac{d+2-\sigma}{(\sigma-d)(2-\lambda)}}.
\end{align}
\end{subequations}
Simulations with $d=1$, $\lambda=0$, and $\gamma=10^{-4}$, are
 characterized by $V\approx 10^2$ and $v_0\approx 10^{-2}$, consistent
 with these scaling laws.

In the complementary case, $\sigma>d+2$, the typical velocity $v_0\sim
1$ is set by the initial conditions because $\bm{\langle}
v^2\bm{\rangle}\sim v_0^2$. Energy balance (\ref{bal}) yields
\begin{equation}
V\sim \gamma^{-\frac{1}{\sigma-d-\lambda}}.
\end{equation}
Simulations with $d=2$, $\lambda=1$, and $\gamma=10^{-2}$ should be
characterized by the injection scale $V\approx 50$, as in this case
$\sigma\cong 4.15$. The data is consistent with this estimate.

As long as the system is sufficiently large, there is no dependence on
the system size (the number of particles). The total energy and the
total dissipation rate are proportional to the system size, and in
general, all thermodynamic properties are extensive.

Based on the theoretical and the simulation results, we conclude that
there may be qualitative differences between the finite energy and
infinite energy cases, but that fundamentally, the steady-state
solutions are of one nature. They represent a nonequilibrium
stationary-state where energy is injected at velocity scale $V$ and
dissipated at velocity scale $v_0$. These scales are set by the
injection rate and the injection protocol.

\section{Time-dependent states}

What happens to these steady states when energy injection is
turned-off?  Steady-state solutions of the type (\ref{powerlaw})
can be realized only up to some upper cutoff. Such truncated
power-law distributions are still compact, and thus, in the
absence of energy input, they should undergo free cooling with all
energy eventually dissipated from the system.

Therefore, we anticipate that there is a time-dependent velocity
cut-off $V(t)$. Below this scale the distribution is nearly the
same as the stationary distribution but above this scale, the
distribution has a sharp tail, analogous to the freely cooling
state (\ref{cooling}). Thus, the distribution is of the form
$f(v,t)\equiv f(v, V)$ such that for $v < V(t)$, $f(v; V) \approx
f_s(v)$ while for $v> V(t)$, the distribution decays faster than a
power law.  Here $f_s(v)$ is the stationary solution of the full
Boltzmann equation.  We assume that the cut-off scale is much
larger than the typical velocity, $V \gg v_0$ and, without loss of
generality, set $v_0 \equiv 1$. The assumption that the
distribution is unmodified below the cut-off velocity is
consistent with the character of the energy cascade. Furthermore,
we expect that the functional form of the cut-off depends only on
the scaled variable, $v/V$.

First, consider the time dependence of the cut-off scale $V(t)$.
Given the assumption that cooling occurs only through a decrease in
the cut-off scale, the rate of change of the energy is
\begin{eqnarray}
\label{dedt}
\frac{dE}{dt}
&=&\frac{d\langle v^2\rangle}{dt}
= \frac{d}{dt}\int^{V(t)} dv\, v^{d+1} f_s(v)\\
&\sim& V^{d+1-\sigma} \frac{dV}{dt}.\nonumber
\end{eqnarray}
The decrease in energy equals the dissipation rate $\Gamma \sim
V^{d+2+\lambda-\sigma}$ from Eq.\ (\ref{dissipation}), showing that
the cut-off velocity obeys Haff's cooling law,
\begin{eqnarray}
\label{Vhaff}
\frac{dV}{dt} = -cV^{1+\lambda}.
\end{eqnarray}
Therefore, the cut-off velocity decays with time as follows
\begin{equation}
\label{decaysoln}
V(t)= \left[\frac{V^\lambda(0)}{1+ c\lambda
V^\lambda(0)t}\right]^{1/\lambda}
\end{equation}
where $V(0)$ is the initial value of the cut-off.

Restricting our attention to one-dimension we seek similarity
solutions of the type
\begin{equation}
\label{scaling}
f(v,t)\simeq f_s(v)\phi\left(\frac{v}{V}\right).
\end{equation}
Here, $f_s(v)$ is the stationary solution of Eq.~(\ref{be-1d}) that
decays as a power-law at large velocities $f_s(v)\simeq
Av^{-2-\lambda}$. The cut-off function approaches unity at small
arguments, $\phi(x)\to 1$ as $x\to 0$ so that the stationary solution
is recovered for $v \ll V$.

Substituting the time dependent form (\ref{scaling}) into the {\it
linear} governing equation (\ref{lbe-1d}) yields
\begin{equation}
-\frac{dV/dt}{V^{1+\lambda}}\phi'(x)=x^{\lambda-1}
\left[p\phi\left(\frac{x}{p}\right)
+q\phi\left(\frac{x}{q}\right)-\phi(x)\right].\quad
\end{equation}
Assuming the cut-off velocity satisfies Eq.~(\ref{Vhaff}) with
constant of proportionality $c$, the scaling function satisfies the
linear and non-local differential equation
\begin{equation}
\label{scaling-eq}
c\,\phi'(x)=x^{\lambda-1}
\left[p\phi\left(\frac{x}{p}\right)
+q\phi\left(\frac{x}{q}\right)-\phi(x)\right],
\end{equation}
with the boundary conditions $\phi(0)=1$ and $\phi(x)\to 0$ as $x \to
\infty$.  Note that $\phi(x)$ must be non-analytic at $x=0$ because
all its derivatives vanish at $x=0$ since $p+q=1$ and $\phi(0)=1$.

For large arguments, the last term on the right hand side dominates,
and therefore, the tail of the distribution is a stretched exponential
as in (\ref{psi})
\begin{equation}
\phi(x)\sim \exp\left(-C\,x^\lambda\right)
\end{equation}
with $C=(\lambda c)^{-1}$ for $\lambda>0$. In the limiting case
$\lambda=0$ all terms on the left-hand side are comparable and the
tail is algebraic: $\phi(x) \sim x^{-\sigma}$ with
\hbox{$c\,\sigma=1-p^{\sigma+1}-q^{\sigma+1}$}. Thus, both the decay
of the cut-off velocity and the tail behavior are as for ordinary
freely cooling solutions, Eq.~(\ref{cooling}).  There is, however, a
difference since the distributions considered here have {\em two}
characteristic velocities, $V$ and $v_0$ and it is only the upper
cut-off, $V$ that evolves in time.  After $V$ and $v_0$ become
comparable, the behavior crosses over to the homogeneous cooling state
\cite{ep,bds} with a single characteristic velocity, $v_0$.

We now focus on completely inelastic hard-spheres ($\lambda=1$ and
$p=q=1/2$) for which an exact solution is possible. Integrating
Eq.~(\ref{scaling-eq}) and imposing $\phi(0)=1$ gives
$c=(1-p^2-q^2)\int_0^\infty dx\, \phi(x)$, but since the cut-off scale
$V$ is defined up to a constant, we may set the integral value,
$\int_0^\infty dx\, \phi(x)=1$ leading to $c=1-p^2-q^2$. When
$p=q=1/2$ then $c=1/2$ and Eq.~(\ref{scaling-eq}) becomes
\begin{equation}
\label{phix-eq}
\phi'(x)=2\left[\phi(2x)-\phi(x)\right].
\end{equation}
This equation can be solved using the Laplace transform
$h(s)=\int dx\, e^{-sx}\, \phi(x)$, that satisfies
the non-local equation
\begin{equation}
\label{hs-eq} (2+s)h(s)=1+h(s/2)
\end{equation}
with the boundary condition $h(0)=1$ set by the normalization.
Since $\phi(x)\to 1$ as $x\to 0$ then $h(s)\to s^{-1}$ as
$s\to\infty$, so we make the transformation
\hbox{$h(s)=s^{-1}\left[1-g(s)\right]$} with $g(0)=1$ and
$g'(0)=-1$. The auxiliary function $g(s)$ satisfies a
recursion-like equation $g(s)=\left(1+s/2\right)^{-1}\,g(s/2)$.
Solving iteratively and invoking $g(0)=1$, the solution is the
infinite product $g(s)=\prod_{n=1}^\infty (1+s/2^{n})^{-1}$, and
the Laplace transform is
\begin{equation}
\label{hs}
h(s)=\frac{1}{s}\left(1-\prod_{n=1}^\infty\frac{1}{1+\frac{s}{2^n}}\right).
\end{equation}
Since the infinite product has a series of simple poles at
\hbox{$s=-2^{-n}$} for every integer $n\geq 1$, the scaling function
is a sum of exponentials
\begin{subequations}
\label{phi-sol}
\begin{align}
\label{phix}
\phi(x)&=\sum_{n=1}^{\infty} a_n \exp\left(-2^n x\right)\\
\label{an}
a_n&=\prod_{\substack{k=1\\k\neq n}}^\infty\frac{1}{1-2^{n-k}},
\end{align}
\end{subequations}
with the coefficients obtained as the residues to the poles
\hbox{$a_n=\lim_{s\to -2^n} [(s+2^n)h(s)]$}. In contrast with
freely cooling states (\ref{cooling}), the scaling function
$\phi(x)$ can be obtained exactly.

\begin{figure}[t]
\includegraphics[width=0.45\textwidth]{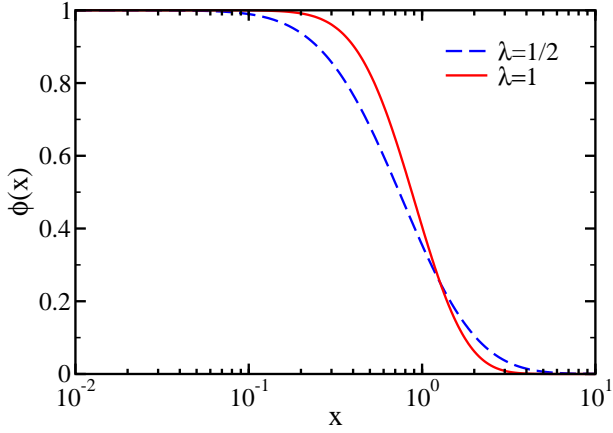}
\caption{The scaling function $\phi(x)$ versus $x$ for $\lambda=1/2$
(dashed line) and $\lambda=1$ (solid line).}
\label{fig:phix}
\end{figure}

The Laplace transform conveniently yields the limiting behaviors of
the scaling function. The simple pole closest to the origin reflects
the tail behavior
\begin{equation}
\phi(x)\simeq a_1\exp(-2\,x)
\end{equation}
as $x\to\infty$ with $a_1=3.46275$ obtained from Eq.~(\ref{an}).  More
interesting is the small $x$ behavior, reflected by the large $s$
behavior
\begin{eqnarray*}
\int_0^\infty dx \left[1-\phi(x)\right]e^{-sx}=s^{-1}g(s)\to
s^{-1}\exp\left[-C\,(\ln s)^2\right]
\end{eqnarray*}
as $s\to \infty$ with $C=(2\ln 2)^{-1}$. The function $g(s)$ was
estimated by replacing the infinite product with a finite product
\begin{equation}
\prod_{n=1}^\infty \frac{1}{1+\frac{s}{2^n}}\cong
\prod_{n=1}^{n_*} \frac{2^n}{s}\to \exp\left[-C\,(\ln s)^2\right]
\end{equation}
with $n_*=\ln_2 s$. Inverting the log-normal Laplace transform using
the steepest descent method, the leading correction to the scaling
function is log-normal as well
\begin{equation}
1-\phi(x)\sim \exp\left[-A\left(\ln \frac{1}{x}\right)^2\right],
\end{equation}
as $x\to 0$ with $A=C/4=(8\ln 2)^{-1}$. Thus, the scaling function is
perfectly flat near the origin as all its derivatives vanish at $x=0$
(figure \ref{fig:phix}). Physically, the small $x$ behavior shows that
there is a sizable range of velocities for which the time-dependent
velocity distribution (\ref{scaling}) coincides with the steady-state
solution, $f(v,t)\simeq f_s(v)$.

\begin{figure}[t]
\includegraphics[width=0.45\textwidth]{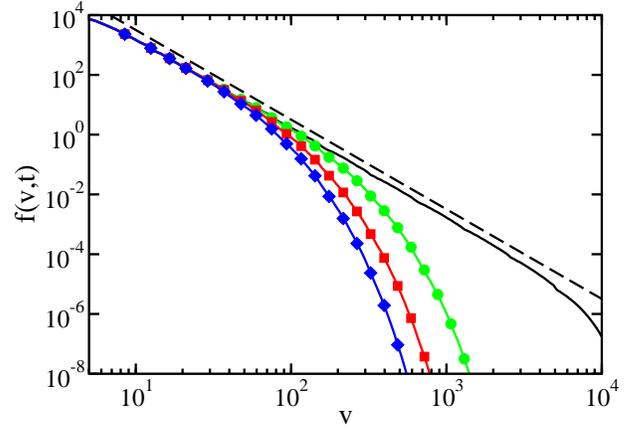}
\caption{The velocity distribution $f(v,t)$ versus $v$. Shown is the
steady-state distribution before the injection is turned-off (solid
line) and at three consecutive and equally-spaced later times
(circles, squares, diamonds) during free cooling.  Also shown for
reference is a dashed line with slope -3. The velocity is in arbitrary
units.}
\label{fig:snapshots}
\end{figure}

The series solution (\ref{phi-sol}) can be straightforwardly
generalized to all $\lambda>0$
\begin{subequations}
\label{phig-sol}
\begin{align}
\label{phix-g}
\phi(x)&=\sum_{n=1}^{\infty} a_n \exp\bigl[-(2^n x)^\lambda\bigr],\\
\label{an-g}
a_n&=\prod_{\substack{k=1\\k\neq n}}^\infty\frac{1}{1-2^{\lambda(n-k)}}.
\end{align}
\end{subequations}
Making the transformation $y=x^\lambda$ and setting the
proportionality constant $c=\lambda^{-1}2^{-\lambda}$ such that
$c=\left(1-2^{-\lambda}\right)\int_0^\infty dy\,\phi(y)$,
Eq.~(\ref{phix-eq}) is generalized,
\hbox{$\phi'(y)=2^\lambda\left[\phi(2^\lambda y)-\phi(y)\right]$}.
Consequently, the Laplace transform is obtained from
Eq.~(\ref{hs}) by replacing $2^n$ with $2^{\lambda n}$, and
repeating the steps leading to (\ref{phi-sol}) gives
(\ref{phig-sol}). Figure \ref{fig:phix} shows the scaling function
for $\lambda=1/2$.  As $\lambda$ decreases the cut-off becomes
less sharp and the flat region near $x=0$ less broad.

In summary, we find that there are time-dependent states associated
with the stationary states. In these transient states, the velocity
distribution is characterized by a cut-off velocity scale that decays
with time according to Haff's law. Below this velocity, the energy
cascade is unaffected and the velocity distribution agrees with the
stationary distribution but above this scale, the distribution is
exponentially suppressed. We relied only on the linear Boltzmann
equation to derive a scaling form for the cut-off function. Of course,
the full nonlinear equation (\ref{be-1d}) is still relevant as it
governs the dynamics of small velocities via the stationary solution
$f_s(v)$. This guarantees that the velocity distribution is properly
normalized, and specifically, that the integral over small velocities
remains finite.

We numerically integrated the hard-sphere Boltzmann equation in
one-dimension to verify these predictions.  Velocity bins are kept,
each with a double precision number representing the number of
particles within that velocity range. In the simulation, two velocity
bins are chosen randomly with a rate proportional to the collision
rate. When two bins ``collide'', particles are transferred from each
into target bins, determined by the collision rule (\ref{rule-1d}).

We generated the stationary distribution by injecting energy at a
fixed rate.  This was done by uniformly removing particles from the
distribution and re-injecting them according to a Gaussian
distribution with a large characteristic velocity.  Once the system
reaches the stationary state, we turn off the energy injection and
observe the distribution $f(v,t)$ as it cools.

\begin{figure}[t]
\includegraphics[width=0.45\textwidth]{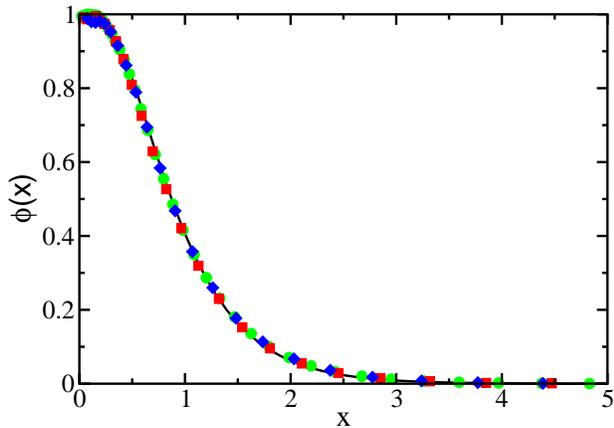}
\caption{The scaling function underlying the velocity
distribution. The velocity distributions in figure
\ref{fig:snapshots} were normalized by the stationary distribution
as in Eq.~(\ref{scaling}). The solid line is the theoretical
scaling function (\ref{phi-sol}).} \label{fig:scaling}
\end{figure}

Figure \ref{fig:snapshots} shows the driven steady-state
distribution
 and the freely cooling distributions at three later times.  The
 results verify that the steady-state has a power-law tail,
 $f_s(v)\sim v^{-3}$ and that the freely cooling distributions
 are close to the steady-state distribution for sufficiently small
 velocities. Figure \ref{fig:scaling} shows
 the same three time-dependent distributions divided by the steady
 distribution as in Eq.~(\ref{scaling}) and rescaled by the cut-off
 velocity $V(t)$ to collapse the data onto the theoretical prediction
 (\ref{phi-sol}).

\begin{figure}[t]
\includegraphics[width=0.45\textwidth]{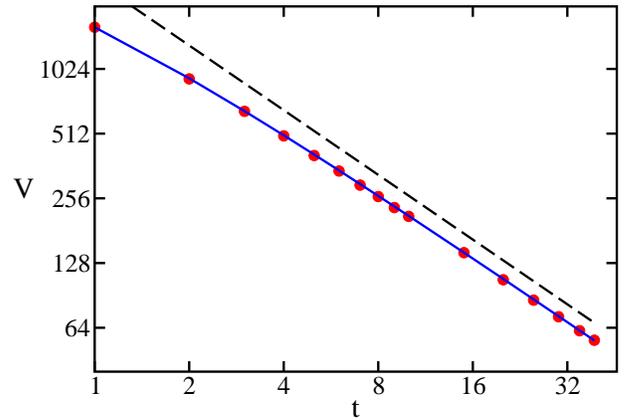}
\caption{The cut-off velocity $V(t)$ as a function of time $t$
(circles).  The solid line is a fit to Eq.~(\ref{decaysoln}).  Also
shown is a broken line with slope $-1$. }
\label{fig:vt}
\end{figure}

The time dependence of the cut-off velocity, given by
Eq.~(\ref{Vhaff}), holds until $V$ is order $v_0\equiv 1$.  Thus,
the lifetime of the collapsing power-law solution approaches a
constant of order unity as $V(0)$ becomes infinite.  During most
of the time that the power-law is collapsing, $V$ decays
algebraically with time, $V(t) \sim t^{-1/\lambda}$.  Figure
\ref{fig:vt} shows the cut-off velocity versus time together with
a fit to the form (\ref{decaysoln}) with $\lambda=1$.  We also
checked that the tail of the cooling distribution is exponential.
We conclude that for completely inelastic hard-spheres, the
simulation results are in excellent agreement with the theoretical
predictions.

\section{Conclusions}

In summary, we find a new class of steady-state and time-dependent
states for inelastic gases. In the nonequilibrium steady-states,
energy is injected at large velocities, it cascades down to small
velocities, and it is dissipated over a power-law range.  Generically,
the steady-state distributions have a power-law high-energy tail. The
characteristic exponents were obtained analytically and they vary with
the spatial dimension and the collision rules. Formally, the
stationary solutions are characterized by an infinite dissipation
rate, while the energy density may be either finite or infinite. In an
actual particle system, these steady-states may be realized only up to
the energy injection scale, so that all thermodynamic characteristics
including the dissipation rate and the energy density are finite.

When injection is turned-off, the velocity distribution is
stationary only in a shrinking range of velocities and it decays
sharply as a stretched exponential at large velocities.  For
completely inelastic collisions, the scaling function underlying
this behavior can be obtained exactly from the linearized
Boltzmann equation and at small velocities, there is a subtle
log-normal correction to the power-law behavior. Although we
analyzed only the one-dimensional case, we expect the same
behavior in higher-dimension. These time-dependent states can be
loosely thought of as a hybrid between a steady-state solution and
the well-known, freely-cooling solution.  Both the time-dependence
of the characteristic velocity and the decay at large velocities
are similar, but not identical, to those occurring for freely
cooling granular gases. After the cut-off velocity becomes
comparable to the typical velocity, the velocity distribution
presumably crosses over to the freely cooling solutions.

Cascade processes occur in many other physical
systems. Mathematically, the inelastic cascade process for
one-dimensional hard-spheres is identical to that found for the
grinding process in Ref.~\cite{bk00}, and in both problems
$\sigma=3$. Indeed, the cascade process (\ref{cp-1d}) is equivalent to
a fragmentation process. In fluid turbulence, the fluid is forced at a
large spatial scale, energy cascades from large scales to small
scales, where it is dissipated due to viscosity \cite{uf}. Actually,
the situation found here for granular gases is analogous to wave
turbulence, that is described by a kinetic theory for wave collisions
\cite{zlf}. One difference with the Kolmogorov spectra of fluid
turbulence is that the characteristic exponents are irrational because
they do not follow from dimensional analysis.

Inelastic cascades are a direct consequence of the collision rule
and they are described by a linear equation. This equation should
be valid under very broad conditions and it can be generalized to
nonuniform distributions of impact angles and collision parameters
as well as polydisperse granular media. Additionally, the cascade
dynamics should extend to viscoelastic collision rules because the
restitution coefficient depends only weakly on the relative
velocity for energetic collisions \cite{bp}.

The most significant condition for these steady-states concerns
the driving mechanism: the injection rate must be small compared
to the collision rate and the injection energy large compared to
the typical energy. We propose that stationary distributions may
be achieved in driven granular gas experiments where energy is
injected at very large velocity scales. Algebraic tails with
exponents comparable with these reported here were observed
recently in sheared granular layers, but these experiments
involved frictional, rather than collisional, dynamics \cite{mn}.

Inelastic cascades should also arise when an energetic particle
hits a static medium of inelastic particles, or alternatively, a
background of slowly moving particles. Indeed, the collision
dynamics in this case reduce to the inelastic cascade discussed in
this paper. This setup may be interesting to study
theoretically and experimentally.

In closing, the kinetic theory of inelastic collisions is
remarkable as the nonlinear Boltzmann equation admits a number of
distinct solutions including steady-states, transient states, and
hybrid states that interpolate between the two.  Nonlinearity,
nonlocality, and the lack of energy conservation are responsible
for this remarkable complexity. We end with an open question: do
other families of solutions exist?

\acknowledgments

We thank P.~L.~Krapivsky, N.~Menon, and V.~Zakharov for useful
discussions. We acknowledge DOE W-7405-ENG-36, NSF DMR-0242402 and
NSF PHY99-07949 for support of this work.

\end{document}